\documentclass[amsmath,amssymb,aps]{revtex4-1}
\usepackage{graphicx}
\usepackage{dcolumn}
\usepackage{bm}
\begin{document}

\title{Quantization of surface plasmon polariton on the metal slab by Green's tensor
method in amplifying and attenuating media }
\author{Z.Allameh}
\affiliation{Institute of Laser and Plasma, Shahid Beheshti University, Tehran, Iran\\}
 \author{R.Roknizadeh}
 \affiliation{
 Department of Physics, Quantum Optics Group\\University of Isfahan, Isfahan, Iran}
\author{R.Masoudi}
\affiliation{
 Institute of Laser and Plasma, Shahid Beheshti University, Tehran, Iran\\}
 
\date{\today}

\begin{abstract}
A quantized form of Surface Plasmon Polariton (SPP) modes propagating  on the metal thin film is provided,  which is based on the Green's tensor method. Since the media will be considered lossy and dispersive, the amplification and attenuation of the SPP modes in various dielectric media, by applying different field frequencies,  can be studied. We will also illustrate the difference between behavior of  coherent and squeezed SPP modes in the amplifying media. 
\end{abstract}
\maketitle

\section{introduction}
The study on surface plasmon polariton \cite{01} is an growing area which has attracted much interest for various applications \cite{002}. Because of the quantum nature of SPP \cite{2,3,4,5}, its applications in some areas such as  quantum information process becomes an active field \cite{02,03,04}. In order to applying the quantum plasmonic, a suitable quantized form of SPP must be provided. In Ref. \cite{05} a quantum mechanical form of SPP's field vectors based on Hopfield theory presented  but in this formalism the dissipation is not considered. Recently in \cite{06} for SPP propagating  in the semi infinite geometry,  we have  proposed another method for quantization based on Green's tensor method \cite{07,08,09,010} which contains the loss. Moreover, this method has the  potential for generalization to the dispersive and inhomogeneous media with different geometries.

In the present contribution  we extend the   technique developed in \cite{06} for quantization of SPP mode for a thin film. It also enable us to studying the SPP phenomena in quantum approach such as amplification or attenuation of the SPP modes for some  quantum (coherent and squeezed SPP) states. 

This paper is structured as follows: the main foundations of quantization of EM fields is provided in section 2. By considering the slab geometry, the procedure of evaluating of the corresponding Green tensor and applying it for obtaining the quantization form of SPP field vectors are presented in section 3. In this section we also investigate some well known relations such as field fluctuations, canonical commutation relations and Langevin equation. In section 4, by applying the quantized form of SPP field, we investigate the influence of the frequency and dielectric media with different optical parameters on the amplification and attenuation of SPP modes. Moreover, we illustrate schematically the difference between behavior of the two modes of SPP( symmetric and antisymmetric) under considered conditions. We also compare the behavior of two modes for two types of states that is possible only in the quantum scheme. A conclusion is given in section 5. 

 \section{Preliminaries}
 The fundamental concepts that form the basis of the quantization procedure by Green’s tensor method are discussed comprehensively in \cite{011,012,013}, so a short  review is presented in this section.
 
  The EM-wave propagates in a dielectric medium with dielectric function $ \epsilon(r,\omega) $, which is related to the complex refractive index
 \begin{equation}
\epsilon(r,\omega)=[n(r,\omega)]^{2}=(\eta(r,\omega)+i\kappa(r,\omega))^{2},\label{1}
\end{equation}
here $ \eta(r,\omega) $ and $\kappa(r,\omega)$ are real and imaginary part of refractive index, respectively. In general, in a range of frequencies, if $\kappa(r,\omega)$ is negative, the dielectric media attenuates the EM- waves,  otherwise  it can be considered as amplifying media.
In order to investigating the behavior of EM- waves propagating  in the dielectric media, in the quantum approach, it is useful to consider the electric and magnetic operators $ (\hat{E}(r,\omega), \hat{B}(r,\omega)) $ according to the vector potential operators $ \hat{A}(r,\omega) $
\begin{align}
\hat{E}(r,\omega)&=\dfrac{\partial\hat{A}(r,\omega)}{\partial {\rm t}},\nonumber\\
\hat{B}(r,\omega)&=\nabla\times\hat{A}(r,\omega).\label{2}
\end{align}
On the other hand, in the frequency domain, the field operators can be considered as positive and negative components:
\begin{equation}
 \hat{E}(r,\omega)= \hat{E}^{+}(r,\omega)+ \hat{E}^{-}(r,\omega),\label{3}
 \end{equation}
where
\begin{equation}
\hat{E}^{\pm}(r,t)=\dfrac{1}{\sqrt{2\pi}}\int_{0}^{+\infty}\text{d}\omega \hat{E}^{\pm}(r,\omega)\exp(\mp i\omega t).\label{4}
\end{equation}
Accordingly for $ \hat{B}(r,\omega) $ and $ \hat{A}(r,\omega) $ we have the similar relations.  By  substituting  the Eqs. \eqref{3} and \eqref{2} into the quantized Maxwell equation \cite{014,015}, and decompose it into  the positive and negative parts,  a general equation for vector potential operator will be obtained, 
\begin{equation}
-\nabla\times\nabla\times \hat{A}^{+}(r,\omega)+\dfrac{\omega^{2}}{c^{2}}\epsilon(r,\omega)\hat{A}^{+}(r,\omega)=-\mu_{0}\hat{j}_{N}^{+}(r,\omega),\label{5}
\end{equation}
where $ \hat{j}_{N}^{+}(r,\omega) $ is the noise current operator associated with the noise sources in the absorbing   
 (or dissipative) media, which is deduced from the fluctuation-dissipation theorem \cite{011,016}. It is convenient to express $ \hat{j}_{N}^{+}(r,\omega) $ according to the normalized noise operator $ \hat{f}(r,\omega) $ 
\begin{equation*}
\hat{j}_{N}^{(+)}(r,\omega)=\sqrt{\alpha(\omega)}\hat{f}(r,\omega),
\end{equation*}
where the coefficient $ \alpha(\omega) $ depends on the optical properties of the media, which satisfy the commutation relations, \begin{align}
[\hat{f}(r,\omega), \hat{f}^{\dagger}(r^{'},\omega^{'})]&=\delta(r-r^{'})\delta(\omega -\omega^{'}),\nonumber\\
[\hat{f}(r,\omega), \hat{f}(r^{'},\omega^{'})]&=[\hat{f}^{\dagger}(r,\omega), \hat{f}^{\dagger}(r^{'},\omega^{'})]=0.\label{5a}
\end{align} 
One of the solution of  Eq.\eqref{5} is based on the standard Green's tensor method
\begin{equation}
\hat{A}^{+}(r,\omega)=-\mu_{0}\int_{-\infty}^{+\infty}\text{d}r^{'}G(r,r^{'},\omega).\hat{j}_{N}^{+}(r^{'},\omega).\label{6}
\end{equation} 
 The Green's tensor must satisfy the Eq. \eqref{5} when the source is replaced by a point source,
\begin{equation}
-\nabla\times\nabla\times G(r,r^{'},\omega)+\dfrac{\omega^{2}}{c^{2}}\epsilon(r,\omega)G(r,r^{'},\omega)=I \delta(r-r^{'}),\label{7}
\end{equation}
where I is a unit tensor. A suitable way for obtaining the system's Green's tensor is eigenmodes expansion method 
 \cite{07,08,09,010}.  Considering the  eigenmodes and eigenvalues form of Eq.\eqref{5}
\begin{equation}
-\nabla\times\nabla\times A_{n}(r,\omega)+\dfrac{\omega^{2}}{c^{2}}\epsilon(r,\omega)A_{n}(r,\omega)=\epsilon(r,\omega)\lambda_{n}A_{n}(r,\omega),\label{8}
\end{equation}
where $ \lambda_{n} $ and $ A_{n}(r,\omega) $ are eigenvalues and eigenmodes, respectively. The eigenmodes satisfy the orthognality condition:
\begin{equation}
\int_{-\infty}^{+\infty}\epsilon(r,\omega) A_{n}(r,\omega)\cdot [A_{m}(r,\omega)]^{*} \text{d}^{3}r=\text{N}_{n}\delta_{nm}.\label{9}
\end{equation}
Therefore the Green's tensor is given by, 
\begin{equation}
G(r,r^{'})=\sum_{n}\dfrac{A_{n}(r) [A_{n}(r^{'})]^{*}}{\text{N}_{n}\lambda_{n}}.\label{10}
\end{equation}
\section{Quantization for SPP field in a metal slab}
\begin{figure}[ht]
\begin{center}
\includegraphics[scale=3]{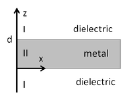}
\end{center}
\caption{Schematic representation of the metal slab embedded between two  dielectrics, SPP propagates on each interfaces of I-II and II-I.}\label{f2}
\end{figure}

In this section we apply the quantization procedure for a metal slab embedded between two dielectrics,  as is shown schematically in Fig.\ref{f2}.
For this system, the dielectric constant can be considered as, 
\begin{align}
\epsilon(r,\omega)=&\epsilon_{d}(\omega)\Theta(-z)+\nonumber\\
&\epsilon_{m}(\omega)\Theta(z)\Theta(-(z-d))+\epsilon_{d}(\omega)\Theta(z-d).\label{11}
\end{align}
In order to obtain the vector potential operator (Eq.\eqref{6}) and quantize it, the  Green's tensor would be derived at the first step.
\subsection{construction of Green's tensor for a slab}
By applying the mode expansion method explained in previous section, the Green's tensor can be obtained. By solving the Eq.\eqref{8} for the geometry depicted in Fig\ref{f2}.  one can find two vector potential modes which correspond to two types of SPP modes. They are symmetric and antisymetric modes with the even (lower sign) and odd(upper sign) vector potential function, respectively,
\begin{align}
&A_{k_{x}}(x,z)=(\hat{x}-i\dfrac{k_{x}}{\nu_{0}}\hat{z})e^{\nu_{0}z}e^{ik_{x}x},\qquad\qquad z<0\nonumber\\
&A_{k_{x}}(x,z)=\qquad\qquad\qquad\qquad\qquad\quad\quad 0<z<d\nonumber\\
&A\lbrace(\hat{x}+i\dfrac{k_{x}}{\nu_{m}}\hat{z})e^{-\nu_{m}z} \mp (\hat{x}-i\dfrac{k_{x}}{\nu_{m}}\hat{z})e^{\nu_{m}(z-d)}\rbrace\times e^{ik_{x}x}\nonumber\\
&A_{k_{x}}(x,z)=\mp (\hat{x}+i\dfrac{k_{x}}{\nu_{0}}\hat{z})e^{-\nu_{0}(z-d)}e^{ik_{x}x},\quad z> d\label{12}
\end{align}
where 
\begin{align}
&A=(1\mp\exp(-\nu_{m}d))^{-1}\nonumber\\
&\nu_{0}^{2}=k_{x}^{2}-\dfrac{\epsilon_{d}{\omega ^{2}}}{c^{2}},\nonumber\\
&\nu_{m}^{2}=k_{x}^{2}-\dfrac{\epsilon_{m}{\omega ^{2}}}{c^{2}}.\label{13}
\end{align}
The SPP modes propagate along the $ x $ axis and $ \nu_{0} $ and $ \nu_{m} $ are the decay coeficients along the $ z $ axis for dielectric and metal region, respectively. The upper sign in Eq.\eqref{12} is related to antisymmetric mode and the lower sign corresponds to the symmetric mode. The frequency of the antisymmetric mode is higher than the frequency of the SPP for a single interface  while the symmetric mode's frequency is lower \cite{01}. Furthermore, odd and even mode's frequency satisfy the dispersion relation:
\begin{equation}
e^{\nu_{m}d}=\mp\dfrac{\dfrac{\epsilon_{m}\nu_{0}}{\epsilon_{d}\nu_{m}}-1}{\dfrac{\epsilon_{m}\nu_{0}}{\epsilon_{d}\nu_{m}}+1}\label{13a}
\end{equation}

On the other hand, by inserting the Eq.\eqref{12} in Eqs.\eqref{8} and \eqref{9}, the eigenvalues $ \lambda_{n} $ and the normalization coefficients $ N_{n} $ can be obtained,
\begin{align}
N_{n}(k_{x})=&2\pi\lbrace \frac{\epsilon_{d}}{\nu_{0}}(1+\dfrac{k_{x}^{2}}{\nu_{0}^{2}})+\nonumber\\
&A^{2}\frac{\epsilon_{m}}{\nu_{m}}(1+\dfrac{k_{x}^{2}}{\nu_{m}^{2}})(1-e^{-\nu_{m}d})\mp 2d(1-\dfrac{k_{x}^{2}}{\nu_{m}^{2}})e^{-\nu_{m}d}\rbrace\nonumber\\
=& 2\pi N^{'}_{n}(k_{x})\nonumber\\
\lambda_{n}=&k_{0}^{2}-\dfrac{k_{x}^{2}-\nu_{m}^{2}}{\epsilon_{m}}\label{14}
\end{align}
According to Eqs. \eqref{12}, \eqref{14} and \eqref{10}, the Green's tensor can be written as:
\begin{align}
&G(r,r^{'},\omega)=\int\text{d}k_{x}\dfrac{1}{N_{n}}\times \dfrac{e^{ik_{x}(x-x^{'})}}{k_{0}^{2}-\dfrac{k_{x}^{2}-\nu_{m}^{2}}{\epsilon_{m}}+i0^{+}}\times\nonumber\\
&\lbrace (\hat{x}-i\dfrac{ k_{x}}{\nu_{0}}\hat{z})e^{\nu_{0}z}\Theta(-z)\mp(\hat{x}+i\dfrac{ k_{x}}{\nu_{0}}\hat{z})e^{-\nu_{0}(z-d)}\Theta(z-d)+\nonumber\\
&A[(\hat{x}+i\dfrac{ k_{x}}{\nu_{m}}\hat{z})e^{-\nu_{m}z}\mp(\hat{x}-i\dfrac{ k_{x}}{\nu_{m}}\hat{z})e^{\nu_{m}(z-d)}]\Theta(z)\Theta(d-z)\rbrace\times\nonumber\\
&\lbrace (\hat{x}-i\dfrac{ k_{x}}{\nu_{0}}\hat{z})e^{\nu_{0}z^{'}}\Theta(-z^{'})\mp(\hat{x}+i\dfrac{ k_{x}}{\nu_{0}}\hat{z})e^{-\nu_{0}(z^{'}-d)}\Theta(z^{'}-d)+\nonumber\\
&A[(\hat{x}+i\dfrac{ k_{x}}{\nu_{m}}\hat{z})e^{-\nu_{m}z^{'}}\mp(\hat{x}-i\dfrac{ k_{x}}{\nu_{m}}\hat{z})e^{\nu_{m}(z^{'}-d)}]\Theta(z^{'})\Theta(d-z^{'})\rbrace \label{15}
\end{align}
In order to evaluate the integral by residu theorem, the poles of the denominator must be evaluated. For analytic solution, it is convenient to consider the $ k_{x} $ very close to the roots of the denominator. We assume the $ k_{spp}^{\mp} $ are the roots, the upper (lower) sign is correspond to odd (even) mode. By taylor expansion for $ \nu_{m} $ about the roots, the first order approximation is given by \cite{010}:
\begin{align}
\nu_{m}(k_{x})\simeq\nu_{m}(k^{\mp}_{spp})+(k_{x}-k^{\mp}_{spp})\dfrac{d\nu_{m}}{dk_{x}}\vert_{k_{x}=k^{\mp}_{spp}}\label{16}
\end{align} 
By applying Eq.\eqref{16}, the denaminator of Eq.\eqref{15} yields:
\begin{align}
&\epsilon_{m}k_{0}^{2}+\nu_{m}(k_{x})-k_{spp}^{\mp}\simeq(k_{x}-k^{\mp}_{spp})\times \lbrace- (k_{x}+k^{\mp}_{spp})+\nonumber\\
&(k_{x}-k^{\mp}_{spp})(\dfrac{d\nu_{m}}{dk_{x}}\vert_{k^{\mp}_{spp}})^{2}+2\nu_{m}(k^{\mp}_{spp})(\dfrac{d\nu_{m}}{dk_{x}}\vert_{k^{\mp}_{spp}})\rbrace \label{17}
\end{align}
As can be seen, the poles of the denominator are the zeros indeed. For simplicity in the subsequent calculations, we use $ k_{s}^{\mp} $ instead of $ k_{spp}^{\mp} $. So by some algebra, the Green's tensor can be obtained as:
\begin{align}
&G(r,r^{'},\omega)=-iD e^{ ik_{s}^{\mp}\vert x-x^{'}\vert}\times\nonumber\\
&\lbrace (\hat{x}-i\dfrac{ k_{s}^{\mp}}{\nu_{0}}\hat{z})e^{\nu_{0}z}\Theta(-z)\mp(\hat{x}+i\dfrac{  k_{s}^{\mp}}{\nu_{0}}\hat{z})e^{-\nu_{0}(z-d)}\Theta(z-d)+\nonumber\\
&A[(\hat{x}+i\dfrac{ k_{s}^{\mp}}{\nu_{m}}\hat{z})e^{-\nu_{m}z}\mp(\hat{x}-i\dfrac{  k_{s}^{\mp}}{\nu_{m}}\hat{z})e^{\nu_{m}(z-d)}]\Theta(z)\Theta(d-z)\rbrace\times\nonumber\\
&\lbrace (\hat{x}-i\dfrac{ k_{s}^{\mp}}{\nu_{0}}\hat{z})e^{\nu_{0}z^{'}}\Theta(-z^{'})\mp(\hat{x}+i\dfrac{ k_{s}^{\mp}}{\nu_{0}}\hat{z})e^{-\nu_{0}(z^{'}-d)}\Theta(z^{'}-d)+\nonumber\\
&A[(\hat{x}+i\dfrac{ k_{s}^{\mp}}{\nu_{m}}\hat{z})e^{-\nu_{m}z^{'}}\mp(\hat{x}-i\dfrac{ k_{s}^{\mp}}{\nu_{m}}\hat{z})e^{\nu_{m}(z^{'}-d)}]\Theta(z^{'})\Theta(d-z^{'})\rbrace \label{18}
\end{align}
where
\begin{equation}
D=\dfrac{\epsilon_{m}}{2\lbrace(-N_{n}^{'}( k_{s}^{\mp})(\nu_{m}( k_{s}^{\mp})\dfrac{d\nu_{m}}{dk_{x}}\vert_{ k_{s}^{\mp}}+ k_{s}^{\mp})\rbrace}.\label{19}
\end{equation}
Also the $ \nu_{0}$'s and $ \nu_{m}$'s in Eq.\eqref{18} are evaluated for $ k_{spp}^{\mp} $. Now by using the Green's tensor, we can quntize the vector potential operator on a slab.
\subsection{Field quantization and canonical commutation relation}
Since, in the system at hand, there is three regions (dielectric- metal- dielectric), the corresponding 3 components of noise current  are given by,
\begin{align}
\hat{j}_{N}^{+}(x,z,\omega)=&\hat{j}_{N}^{d+}(x,z,\omega)\Theta(-z)+\hat{j}_{N}^{d+}(x,z,\omega)\Theta(z-d)+\nonumber\\
&\hat{j}_{N}^{m+}(x,z,\omega)\Theta(z)\Theta(d-z),\nonumber\\
=&[\sqrt{\alpha^{d}(\omega)}\Theta(-z)+\sqrt{\alpha^{d}(\omega)}\Theta(z-d)+\nonumber\\
&\sqrt{\alpha^{m}(\omega)}\Theta(z)\Theta(d-z)]\hat{f}(x,z,\omega).\label{20}
\end{align}
By applying the Eqs. \eqref{20}, \eqref{18} and \eqref{6} and some calculations, one can obtain the  vector potential operator and introduce the annihilation and creation operators like the procedure in \ref{06}
\begin{align}
&\hat{A}^{+}(r,\omega)=i\mu_{0}D(\dfrac{\beta^{'}(\omega)}{2k^{\mp}_{sI}})^{\frac{1}{2}}\times\nonumber\\
&\lbrace (\hat{x}-i\dfrac{ k_{s}^{\mp}}{\nu_{0}}\hat{z})e^{\nu_{0}z}\Theta(-z)\mp(\hat{x}+i\dfrac{  k_{s}^{\mp}}{\nu_{0}}\hat{z})e^{-\nu_{0}(z-d)}\Theta(z-d)+\nonumber\\
&A[(\hat{x}+i\dfrac{ k_{s}^{\mp}}{\nu_{m}}\hat{z})e^{-\nu_{m}z}\mp(\hat{x}-i\dfrac{  k_{s}^{\mp}}{\nu_{m}}\hat{z})e^{\nu_{m}(z-d)}]\Theta(z)\Theta(d-z)\rbrace\times\nonumber\\
&\lbrace \hat{a}^{\mp}_{R}(x,\omega)+\hat{a}^{\mp}_{L}(x,\omega)\rbrace , \label{21}
\end{align}\\
here $ k^\mp_{sI}( k^\mp_{sR}) $ is the imaginary (real) part of the $ k_{s}^\mp $ and
\begin{align}
\beta^{'}(\omega)=&\vert\alpha^{d}(\omega)\vert (1+\dfrac{\vert k_{s}^{\mp}\vert^{2}}{\vert \nu_{0}\vert^{2}})\dfrac{2}{\nu_{0}+\nu_{0}^{*}}+\nonumber\\
&\vert\alpha^{m}(\omega)\vert\vert A\vert^{2}\times\lbrace(1+\dfrac{\vert k_{s}^{\mp}\vert^{2}}{\vert \nu_{m}\vert^{2}})\dfrac{2(1-e^{-(\nu_{m}+\nu_{m}^{*})d})}{\nu_{m}+\nu_{m}^{*}}\nonumber\\
&\mp(1-\dfrac{\vert k_{s}^{\mp}\vert^{2}}{\vert \nu_{m}\vert^{2}})\dfrac{2(e^{-\nu_{m}^{*}d}-e^{-\nu_{m}d})}{\nu_{m}-\nu_{m}^{*}} \rbrace .\label{22}
\end{align}
In Eq.\eqref{21} the operators $ \hat{a}^{\mp}_{R}$  and  $\hat{a}^{\mp}_{L} $ indicate the annihilation of the SPP mode with symmetric ($ {+} $) or antisymmetric ($ {-} $) field function that propagates rightwards and leftwards respectively and have the explicit form as:
\begin{align}
&\hat{a}^{\mp}_{R}(x,\omega)=(\dfrac{2k^{\mp}_{sI}}{\beta^{'}(\omega)})^{\frac{1}{2}} \int_{-\infty}^{\infty}\int_{-\infty}^{x}\text{d}x^{'}\text{d}z^{'} e^{ik_{s}^{\mp}( x-x^{'})}\hat{j}_{N}^{+}(x^{'},z^{'},\omega)\cdot\nonumber\\
&\lbrace (\hat{x}-i\dfrac{ k_{s}^{\mp}}{\nu_{0}}\hat{z})e^{\nu_{0}z^{'}}\Theta(-z^{'})\mp(\hat{x}+i\dfrac{  k_{s}^{\mp}}{\nu_{0}}\hat{z})e^{-\nu_{0}(z^{'}-d)}\Theta(z^{'}-d)+\nonumber\\
&A[(\hat{x}+i\dfrac{ k_{s}^{\mp}}{\nu_{m}}\hat{z})e^{-\nu_{m}z^{'}}\mp(\hat{x}-i\dfrac{  k_{s}^{\mp}}{\nu_{m}}\hat{z})e^{\nu_{m}(z^{'}-d)}]\Theta(z^{'})\Theta(d-z^{'})\rbrace  \label{23a}
\end{align}
and
\begin{align}
&\hat{a}^{\mp}_{L}(x,\omega)=(\dfrac{2k^{\mp}_{sI}}{\beta^{'}(\omega)})^{\frac{1}{2}} \int_{-\infty}^{\infty}\int_{x}^{\infty}\text{d}x^{'}\text{d}z^{'} e^{-ik_{s}^{\mp}( x-x^{'})}\hat{j}_{N}^{+}(x^{'},z^{'},\omega)\cdot\nonumber\\
&\lbrace (\hat{x}-i\dfrac{ k_{s}^{\mp}}{\nu_{0}}\hat{z})e^{\nu_{0}z^{'}}\Theta(-z^{'})\mp(\hat{x}+i\dfrac{  k_{s}^{\mp}}{\nu_{0}}\hat{z})e^{-\nu_{0}(z^{'}-d)}\Theta(z^{'}-d)+\nonumber\\
&A[(\hat{x}+i\dfrac{ k_{s}^{\mp}}{\nu_{m}}\hat{z})e^{-\nu_{m}z^{'}}\mp(\hat{x}-i\dfrac{  k_{s}^{\mp}}{\nu_{m}}\hat{z})e^{\nu_{m}(z^{'}-d)}]\Theta(z^{'})\Theta(d-z^{'})\rbrace \label{23}
\end{align}
 By some calculations one can find that the annihilation and creation operators satify the commutation relation,
 \begin{align}
&[\hat{a}^{\mp}_{R}(x,\omega),\hat{a}^{\mp\dagger}_{R}(x^{'},\omega^{'})]=[\hat{a}^{\mp}_{L}(x^{'},\omega^{'}),\hat{a}^{\mp\dagger}_{L}(x,\omega)]=\nonumber\\
&\delta(\omega -\omega^{'})\exp (ik^{\mp}_{sR}(x-x^{'}))\exp( -k^{\mp}_{sI}\vert x-x^{'}\vert ),\label{36}\\
&[\hat{a}^{\mp}_{R}(x,\omega),\hat{a}^{\mp\dagger}_{L}(x^{'},\omega^{'})]=[\hat{a}^{\mp}_{L}(x^{'},\omega^{'}),\hat{a}^{\mp\dagger}_{R}(x,\omega)]=\nonumber\\
&\delta(\omega - \omega^{'})\Theta(x-x^{'})\dfrac{2k^{\mp}_{sI}}{k^{\mp}_{sR}}\exp(-k^{\mp}_{sI}(x-x^{'}))\sin k^{\mp}_{sR}(x-x^{'}).\label{24}
\end{align}
On the other hand, one can explore the canonical commutation relation and obtain, 
\begin{align}
[\hat{A}(r,t), -\epsilon_{0}\hat{E}(r^{'},t)]=\int_{0}^{\infty}\text{d}\omega \dfrac{i\beta^{'}(\omega)}{\pi\epsilon_{0}c^{2}\omega\gamma^{'}(\omega)}\text{Im}G(r,r^{'},\omega), \label{25}
\end{align}   
The details of the calculations and the explicite form of $ \gamma^{'} $ are given in Appendix A. By considering the general property of Green's tensor $ \lim_{\vert \omega\vert\rightarrow\infty}\frac{\omega^{2}}{c^{2}}G(r,r^{'},\omega)=-\delta(r-r^{'}) $ and assuming that:
\begin{equation}
 \beta^{'}(\omega)=2\hbar\epsilon_{0}\omega^{2}\gamma^{'}(\omega)\label{26}
\end{equation}
one can prove that the canonical commutation relation is satisfied. 
\begin{align}
[\hat{A}(r,t), -\epsilon_{0}\hat{E}(r^{'},t)]= i\hbar\delta(r-r^{'}).\label{27}
\end{align}
On the other hand by substituting Eqs.\eqref{22} and \eqref{70} into Eq.\eqref{26} the explicit form of $ \alpha^{m}(\omega) $ and  $ \alpha^{d}(\omega) $ can be obtained. 
\begin{align}
&\vert\alpha^{m}(\omega)\vert =2\hbar \omega^{2}\epsilon_{0}\text{Im}\epsilon_{m}(\omega),\nonumber\\
&\vert\alpha^{d}(\omega)\vert =2\hbar \omega^{2}\epsilon_{0}\text{Im}\epsilon_{d}(\omega).\label{28}
\end{align}
\section{Numerical results: Exploring the amplifyed and attenuated conditions for coherent and squeezed symmetric and antisymmetric SPP modes}
\subsection{magnetic field}
In order to investigate the variation of the SPP's modes propagating in amplifying and attenuating media, we need to calculate the magnetic field. For simplicity, we consider that the SPP waves propagate rightwards. By applying Eqs. \eqref{23a}, \eqref{21} and \eqref{2} magnetic field operator can be obtained:
\begin{align}
&\hat{H}^{+}(r,\omega)=iD(\dfrac{\beta^{'}(\omega)}{2k^{\mp}_{sI}})^{\frac{1}{2}}\times\hat{a}^{\mp}_{R}(x,\omega)\times\nonumber\\
&\lbrace (\nu_{0}-\dfrac{ k_{s}^{\mp 2}}{\nu_{0}})e^{\nu_{0}z}\Theta(-z)\mp(-\nu_{0}+\dfrac{  k_{s}^{\mp 2}}{\nu_{0}})e^{-\nu_{0}(z-d)}\Theta(z-d)+\nonumber\\
&A[(-\nu_{m}+\dfrac{ k_{s}^{\mp 2}}{\nu_{m}})e^{-\nu_{m}z}\mp(\nu_{m}-\dfrac{  k_{s}^{\mp 2}}{\nu_{m}})e^{\nu_{m}(z-d)}]\Theta(z)\Theta(d-z)\rbrace ,\label{29}
\end{align} 
where  $ \hat{a}^{\mp}_{R}(x,\omega) $ satisfy quantum Langevien equation:
\begin{equation}
\dfrac{\partial \hat{a}^{\mp}_{R}(x,\omega)}{\partial x}=ik_{spp}^{\mp} \hat{a}^{\mp}_{R}(x,\omega)+\hat{F}^{\mp}(x,\omega),\label{30}
\end{equation}
where
\begin{align}
&\hat{F}^{\mp}_{R}(x,\omega)=(\dfrac{2k^{\mp}_{sI}}{\beta^{'}(\omega)})^{\frac{1}{2}} \int_{-\infty}^{\infty}\text{d}z^{'} \hat{j}_{N}^{+}(x,z^{'},\omega)\cdot\nonumber\\
&\lbrace (\hat{x}-i\dfrac{ k_{s}^{\mp}}{\nu_{0}}\hat{z})e^{\nu_{0}z^{'}}\Theta(-z^{'})\mp(\hat{x}+i\dfrac{  k_{s}^{\mp}}{\nu_{0}}\hat{z})e^{-\nu_{0}(z^{'}-d)}\Theta(z^{'}-d)+\nonumber\\
&A[(\hat{x}+i\dfrac{ k_{s}^{\mp}}{\nu_{m}}\hat{z})e^{-\nu_{m}z^{'}}\mp(\hat{x}-i\dfrac{  k_{s}^{\mp}}{\nu_{m}}\hat{z})e^{\nu_{m}(z^{'}-d)}]\Theta(z^{'})\Theta(d-z^{'})\rbrace  \label{31}
\end{align}
here $ \hat{F}^{\mp}(x,\omega) $ is the operator associated to the Langevien noise source. On the other hand   by attention to the explicit form of  $ \hat{F}^{\mp}(x,\omega) $   one can find that this operator can  only appear in the absorbing(or dissipative) media when noise sources are exist. \\
 By solving the Eq.\eqref{30} and considering the general property of noise source operators $ \langle \hat{j}_{N}^{+}(x,z,\omega)\rangle =0 $,   the average form of Eq. \eqref{29} is given by
 \begin{align}
&\langle \hat{H}^{+}(r,\omega)\rangle =iD(\dfrac{\beta^{'}(\omega)}{2k^{\mp}_{sI}})^{\frac{1}{2}} \times e^{i k_{s}^{\mp}x}\langle \hat{a}^{\mp}_{R}(\omega)\rangle\times\nonumber\\
&\lbrace (\nu_{0}-\dfrac{ k_{s}^{\mp 2}}{\nu_{0}})e^{\nu_{0}z}\Theta(-z)\mp(-\nu_{0}+\dfrac{  k_{s}^{\mp 2}}{\nu_{0}})e^{-\nu_{0}(z-d)}\Theta(z-d)+\nonumber\\
&A[(-\nu_{m}+\dfrac{ k_{s}^{\mp 2}}{\nu_{m}})e^{-\nu_{m}z}\mp(\nu_{m}-\dfrac{  k_{s}^{\mp 2}} {\nu_{m}})e^{\nu_{m}(z-d)}]\Theta(z)\Theta(d-z)\rbrace . \label{32}
\end{align}
The average  of magnetic field operator can be  considered for  the different kinds of SPP states like coherent and squeezed states \cite{4,5}. The  annihilation and creation operators of the SPP modes obey the relations of bosonic operators. Therefore, when the SPP states are prepared in coherent and squeezed states the following relations can be considered,
\begin{align}
&\hat{a}^{\mp}_{R}(\omega)\vert\alpha\rangle = \alpha\vert\alpha\rangle .\label{33}\\
&\hat{a}^{\mp}_{R}(\omega)\vert \xi , \alpha\rangle =(\mu {\alpha}(\omega)-\nu {\alpha}(\omega))\vert\alpha\rangle, \label{34}
\end{align} 
where $ \alpha =\vert \alpha\vert e^{i\theta} $, $ \mu=\cosh( \vert \xi\vert)$ and $ \nu= \sinh( \vert \xi\vert) e^{i\theta_{\xi}}$.   $ \vert \xi\vert $ and $ \theta_{\xi} $ are the absolute value  and argument of squeezed parameter, respectively. 
 By applying the Eqs.\eqref{33} and \eqref{34}, the magnetic field average can be calculated for coherent and squeezed SPP states.
\subsection{Studying the influence of frequency on the SPP modes}
\begin{figure}[ht]
\begin{center}
\includegraphics[scale=3]{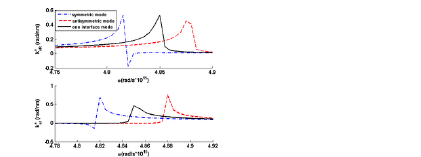}
\end{center}
\caption{Symmetric and antisymmetric SPP modes dispersion relation. (a) real part  and (b) imaginary part of the wave number. Where d=60 nm, the dielectric media with $ n=1.9726-i0.081 $ and the metal film with $ \omega_{p}=14.02\times 10^{15} $(Rad/s) and $ \gamma =6.25\times 10^{13} $(Rad/s) have been considered. (The data are in \cite{017}).}\label{f3}
\end{figure}

The optical properties of the dielectric media adjusted to the metal film can affect the properties of SPPs. According to the intrinsic absorption property of the metal, SPP modes suffer damping which reduces the SPP length propagation. On the other hand the dielectric media with negative imaginary part of refractive index $(n=n_{d}+ik_{d})$ can act as gain media. When the gain of dielectric media is sufficient to compensate the loss in the metals, the $ k_{sI}^{\mp} $ will be negative and the system acts as amplifying media otherwise  for positive $ k_{sI}^{\mp} $  the system can be considered as attenuated media \cite{018,019,020,021,022,023}.\\
The variation of $ k_{sI}^{\mp} $ according to the frequency can be shown by the dispersion relation diagram. It is depicted in figure \ref{f3}.
In Fig. \ref{f3} despite of different behavior for symmetric and antisymmetric modes, there is a same prediction.\\
In general,  in the frequency ranges that in the loss case  $ k_{sI}^{\mp}  $ is negative, the SPP modes can be amplified otherwise they are attenuated.\\
Moreover, it can be shown that for thick film the dispersion relation of symmetric and antisymmetric modes are identical and are in accordance with the one interface SPP mode \cite{018}. 
\subsection{Studying the influence of the dielectric media on the SPP modes}
\begin{figure}[ht]
\begin{center}
\includegraphics[scale=3]{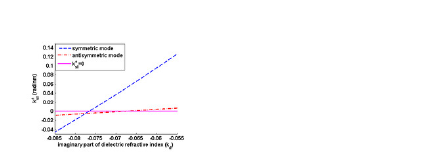}
\end{center}
\caption{Variation of $ k_{sI} $ of the SPP modes in the vicinity of different gain media. where $ n_{d}=1.9726 $ and $ \omega =4.8\times 10^{15} Rad/s$.}\label{f4}
\end{figure}

Besides the frequency, the different gain media (dielectric media) can affect the SPP modes behavior. This is illustrated schematically in Fig. \ref{f4}.\\
Fig.\ref{f4} shows some interesting points: first, any gain media adjusted to the metal thin film can not amplify the SPP modes. Indeed, for amplifying $ k_{d} $ and $ k^{\mp}_{sI} $ must be negative simultaneously. Second, for a particular $ k_{d} $ the operation of the system is different for symmetric and antisymmetric modes. This is discussed in more detail later.
 Third, the slope of the lines indicates that for a given $ k_{d} $ range, the variation of the symmetric mode in comparison with $  k^{\mp}_{sI}=0 $ (no gain and no loss) is very greater than the antisymmetric mode.
 Fourth, except of very small range of  $ k_{d} $ (very close to the $ k^{\mp}_{sI}=0 $ for symmetric mode) the rate of amplifying or attenuating symmetric mode is very faster than the antisymmetric mode.\\
In order of illustration and comparison the variation of SPP modes in these ranges, we plot the magnetic field average Eq. \eqref{32} for coherent symmetric and antisymmetric modes for different ranges of $ k_{d} $ . 
\subsubsection{Investigating the variation of coherent SPP modes under different gain media}
\begin{figure}[ht]
\begin{center}
\includegraphics[scale=3]{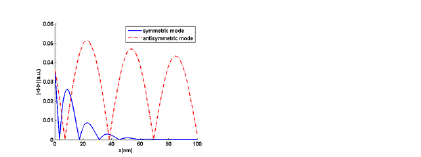}
\end{center}
\caption{Attenuation of two modes on the $ z=0 $ interface. where $ \vert \alpha\vert^{2} =7 $, $ \theta = 1.5  $ Rad.  $ k_{sI}^{+}=7.8029\times 10^{7} $(Rad/nm) and $ k_{sI}^{-}=2.7617\times 10^{6} $(Rad/nm) .}\label{f5}
\end{figure}

\begin{figure}[ht]
\begin{center}
\includegraphics[scale=3]{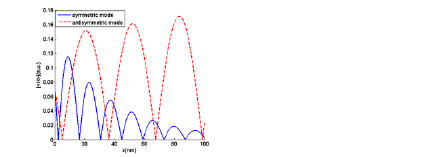}
\end{center}
\caption{Attenuation of symmetric mode$ k_{sI}^{+}=2.5858\times 10^{7} $(Rad/nm) and amplification of antisymmetric mode$ k_{sI}^{-}=-1.96\times 10^{6} $ (Rad/nm) on the $ z=0 $ interface.}\label{f6}
\end{figure}

\begin{figure}[ht]
\begin{center}
\includegraphics[scale=3]{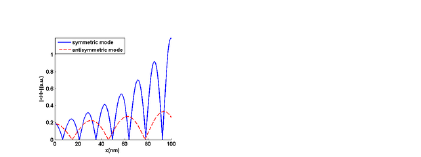}
\end{center}
\caption{Amplification of two modes on the $ z=0 $ interface. where $ k_{sI}^{+}=
 -1.8673\times 10^{7} $(Rad/nm) and $ k_{sI}^{-}= -6.1217\times 10^{6} $ (Rad/nm).}\label{f7}
\end{figure}

\begin{figure}[ht]
\begin{center}
\includegraphics[scale=3]{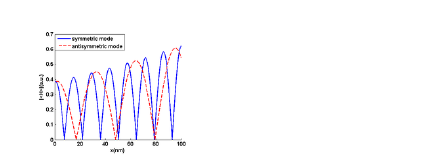}
\end{center}
\caption{The same behavior of two modes on the $ z=0 $ interface. where $ k_{sI}^{+}=k_{sI}^{-}=
 -4.81\times 10^{6} $(Rad/nm) .}\label{f8}
\end{figure}

 According to Fig.\ref{f4}, for different ranges of $ k_{d} $ the SPP modes suffer different conditions. The first condition is where two SPP modes are attenuated. For instance, by considering the dielectric media with $n=0.9726-i0.063 $ the $ k^{\mp}_{sI} $ of the SPP modes are positive. It means that the gain of the dielectric media can not compensate the loss of the metal and the SPP modes are also attenuated. It is shown schematically in Fig.\ref{f5} .
In the second condition, the system's operation is different for symmetric and antisymmetric modes. For example, the dielectric media with $n=0.9726-i0.072 $ cause the system attenuates the symmetric mode but amplifies the antisymmetric mode. It is shown in Fig.\ref{f6}.
The amplification of symmetric and antisymmetric modes simultaneously is occurred in the third condition where  for a given $ k_{d} $,  the $ k^{\mp}_{sI} $  is negative for two modes. Fig. \ref{f7} shows this condition where the refractive index of dielectric media is $n=0.9726-i0.08 $. It means that for the two modes  the gain of the dielectric media is sufficient to overcome the loss of the metal. \\
As the Fig.\ref{f4} illustrates except the small range of  $ k_{d} $,the magnitude of $ k_{sI}^{+}$ is greater than  $ k_{sI}^{-}$, therefore for a special condition the rate of variation for symmetric mode is noticeable than the antisymetric mode, as Figs.(\ref{f5}-\ref{f7}) demonstrate obviously. Moreover by comparison the Figs.(\ref{f5}-\ref{f7}), one can find that for different conditions, the changing of symmetric mode behavior is more noticeable than another.\\
A very interesting point is that, only for one magnitude of $ k_{d} $, the behavior of two modes is the same. It is the point of intersection of two graphs in Fig.\ref{f4}. Fig.\ref{f8} shows this condition where the optical property of dielectric media is  $n=0.9726-i0.07747 $.
 Moreover, the same figures are obtained for the modes that propagate along the $ z=d $ interface.
\subsubsection{Comparison between coherent and squeezed SPP modes state}
\begin{figure}[ht]
\begin{center}
\includegraphics[scale=3]{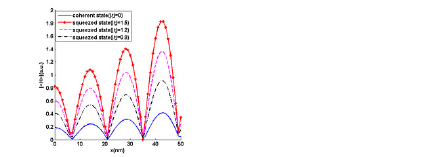}
\end{center}
\caption{The difference between the average of the symmetric mode's magnetic field. where $n=0.9726-i0.08 $ is chosen for dielectric media.}\label{f9}
\end{figure}

\begin{figure}[ht]
\begin{center}
\includegraphics[scale=3]{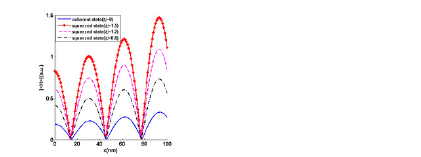}
\end{center}
\caption{The difference between the average of the antisymmetric mode's magnetic field. }\label{f10}
\end{figure}

By applying Eqs.\eqref{33} and \eqref{34} into the Eq.\eqref{32}, the average of magnetic field is obtained for coherent and squeezed SPP state. For more investigation, we consider $ \alpha =\vert\alpha\vert e^{i\theta} $ as a complex number. The influence of the phase $ \theta $ on the average of the SPP magnetic filed for one interface system is studied in \cite{06}. Such result is obtained for two interfaces system which shows that the  difference between coherent and squeezed state is significant for $ \theta =1.5 $. In Fig.\ref{f9}, the difference between the magnetic field average for coherent and squeezed states that propagate in the amplifying system is illustrated.
The average of magnetic field for antisymmetric mode is shown in Fig.\ref{f10}.
From Figs. \ref{f9} and \ref{f10} one can deduce that the drastic difference is occurred between squeezed and coherent states. It can be generalized to other amplifying or attenuating system.
 \section{conclusion}
In this paper we have provided another approach for quantization of SPP modes on the thin film structure, based on Green's tensor method which  contains some new quantum concepts in the SPP field such as noise current, field fluctuations, and Langevin equations. Moreover, this approach enable us to study the influence of the different conditions on the propagation of the SPP modes. The results are classified as follows:\\
First, for certain media, the variation of frequency can cause SPP modes amplification or attenuation. It also has been shown that the behavior of symmetric and antisymmetric modes is different in frequency domain.\\
Second, for certain frequency, the amplifying or attenuating SPP modes is dependent on the optical parameter of the dielectric media adjusted to the metal film. We have also compared the behavior of two SPP modes with each other for different media.\\
Third, we have illustrated that the drastic difference is between different types of SPP modes, i. e.,  coherent and squeezed states.
\section*{appendix A}
In order to prove the Eq.\eqref{25}, it is neccesary to derive some relations:
\begin{align}
&[\hat{A}(r,t), -\epsilon_{0}\hat{E}(r^{'},t)]=\int_{0}^{\infty}\text{d}\omega \vert D\vert^{2}\mu_{0}^{2}\dfrac{i\omega\epsilon_{0}\beta^{'}(\omega)}{2\pi k^{\mp}_{sI}}\times\nonumber\\
&\dfrac{\vert k_{s}^{\mp}\vert^{2}}{k^{\mp}_{sR}}\lbrace\dfrac{e^{ik_{s}^{\mp}\vert x-x^{'}\vert}}{k_{s}^{\mp}}+\dfrac{e^{-ik_{s}^{{*\mp}}\vert x-x^{'}\vert}}{k^{*\mp}_{s}}\rbrace\times\nonumber\\
&\lbrace (\hat{x}-i\dfrac{ k_{s}^{\mp}}{\nu_{0}}\hat{z})e^{\nu_{0}z}\Theta(-z)\mp(\hat{x}+i\dfrac{  k_{s}^{\mp}}{\nu_{0}}\hat{z})e^{-\nu_{0}(z-d)}\Theta(z-d)+\nonumber\\
&A[(\hat{x}+i\dfrac{ k_{s}^{\mp}}{\nu_{m}}\hat{z})e^{-\nu_{m}z}\mp(\hat{x}-i\dfrac{  k_{s}^{\mp}}{\nu_{m}}\hat{z})e^{\nu_{m}(z-d)}]\Theta(z)\Theta(d-z)\rbrace\times\nonumber\\
&\lbrace (\hat{x}+i\dfrac{ k_{s}^{{*\mp}}}{\nu^{*}_{0}}\hat{z})e^{\nu^{*}_{0}z^{'}}\Theta(-z^{'})\mp(\hat{x}-i\dfrac{  k_{s}^{{*\mp}}}{\nu^{*}_{0}}\hat{z})e^{-\nu^{*}_{0}(z^{'}-d)}\Theta(z^{'}-d)+\nonumber\\
&A^{*}[(\hat{x}-i\dfrac{ k_{s}^{{*\mp}}}{\nu^{*}_{m}}\hat{z})e^{-\nu^{*}_{m}z^{'}}\mp(\hat{x}+i\dfrac{k_{s}^{{*\mp}}}{\nu^{*}_{m}}\hat{z})e^{\nu^{*}_{m}(z^{'}-d)}]\Theta(z^{'})\Theta(d-z^{'})\rbrace .\label{68}
\end{align}
On the other hand, the Green's tensor for a dielectric-metal- dielectric structure (see Eq.\eqref{18}) satisfy the following relation:
\begin{align}
&\int \text{d}s \text{Im}\epsilon(s,\omega)G(r,s,\omega) \cdot G^{*}(s,r^{'},\omega)=\nonumber\\
&\dfrac{\vert  k_{s}^{\mp}\vert^{2}}{2k^{\mp}_{sR}k^{\mp}_{sI}}\gamma^{'}(\omega) \vert D\vert^{2}\lbrace\dfrac{e^{ik_{s}^{\mp}\vert x-x^{'}\vert}}{k_{s}^{\mp}}+\dfrac{e^{-ik^{*\mp}_{s}\vert x-x^{'}\vert}}{k^{*\mp}_{s}}\rbrace\times\nonumber\\
&\lbrace (\hat{x}-i\dfrac{ k_{s}^{\mp}}{\nu_{0}}\hat{z})e^{\nu_{0}z}\Theta(-z)\mp(\hat{x}+i\dfrac{  k_{s}^{\mp}}{\nu_{0}}\hat{z})e^{-\nu_{0}(z-d)}\Theta(z-d)+\nonumber\\
&A[(\hat{x}+i\dfrac{ k_{s}^{\mp}}{\nu_{m}}\hat{z})e^{-\nu_{m}z}\mp(\hat{x}-i\dfrac{  k_{s}^{\mp}}{\nu_{m}}\hat{z})e^{\nu_{m}(z-d)}]\Theta(z)\Theta(d-z)\rbrace\times\nonumber\\
&\lbrace (\hat{x}+i\dfrac{ k_{s}^{{*\mp}}}{\nu^{*}_{0}}\hat{z})e^{\nu^{*}_{0}z^{'}}\Theta(-z^{'})\mp(\hat{x}-i\dfrac{  k_{s}^{{*\mp}}}{\nu^{*}_{0}}\hat{z})e^{-\nu^{*}_{0}(z^{'}-d)}\Theta(z^{'}-d)+\nonumber\\
&A^{*}[(\hat{x}-i\dfrac{ k_{s}^{{*\mp}}}{\nu^{*}_{m}}\hat{z})e^{-\nu^{*}_{m}z^{'}}\mp(\hat{x}+i\dfrac{k_{s}^{{*\mp}}}{\nu^{*}_{m}}\hat{z})e^{\nu^{*}_{m}(z^{'}-d)}]\Theta(z^{'})\Theta(d-z^{'})\rbrace ,\label{69}
\end{align}
here
\begin{align}
\gamma^{'}(\omega)=&\text{Im}\epsilon_{m} (1+\dfrac{\vert k_{s}^{\mp}\vert^{2}}{\vert \nu_{0}\vert^{2}})\dfrac{2}{\nu_{0}+\nu_{0}^{*}}+\nonumber\\
&\text{Im}\epsilon_{d}\vert A\vert^{2}\times\lbrace(1+\dfrac{\vert k_{s}^{\mp}\vert^{2}}{\vert \nu_{m}\vert^{2}})\dfrac{2(1-e^{-(\nu_{m}+\nu_{m}^{*})d})}{\nu_{m}+\nu_{m}^{*}}\nonumber\\
&\mp(1-\dfrac{\vert k_{s}^{\mp}\vert^{2}}{\vert \nu_{m}\vert^{2}})\dfrac{2(e^{-\nu_{m}^{*}d}-e^{-\nu_{m}d})}{\nu_{m}-\nu_{m}^{*}} \rbrace .\label{70}
\end{align}
and  $ \epsilon(s,\omega) $ has been given in Eq.\eqref{11}. By rewritting Eq.\eqref{68} according to Eq.\eqref{69} and considering the general property of Green's tensor\cite{024}
\begin{equation}
\int \text{d}s \text{Im}\epsilon(s,\omega)G(r,s,\omega) \cdot G^{*}(s,r^{'},\omega)=\dfrac{c^{2}}{\omega^{2}}\text{Im}G(r,r^{'},\omega)\label{64}
\end{equation} 
The desired  equation can be deduced:
\begin{align*}
&[\hat{A}(r,t), -\epsilon_{0}\hat{E}(r^{'},t)]=\int \text{d}\omega \dfrac{i\omega\epsilon_{0}\mu_{0}^{2}\beta^{'}(\omega)}{\pi \gamma^{'}(\omega)}\dfrac{c^{2}}{\omega^{2}}\text{Im}G(r,r^{'},\omega).
\end{align*} 

\end{document}